\documentclass[jgrga]{agu2001} 

\usepackage{graphicx}
\newcommand{\be}{\begin{equation}}
\newcommand{\ee}{\end{equation}}
\newcommand{\ba}{\begin{eqnarray}}
\newcommand{\ea}{\end{eqnarray}}
\authorrunninghead{SORNETTE and WERNER}

\titlerunninghead{CONSTRAINTS ON THE SMALLEST TRIGGERING EARTHQUAKE}

\authoraddr{D. Sornette, Department of Earth and Space Sciences, and
   Institute of Geophysics and Planetary Physics, University of
   California, Los Angeles, California 90095, USA and Laboratoire de
   Physique de la Mati\`ere Condens\'ee, CNRS UMR6622, Universit\'e de
   Nice-Sophia Antipolis, Parc Valrose, 06108 Nice Cedex 2,
   France. (sornette@moho.ess.ucla.edu)}

\authoraddr{M. J. Werner, Department of Earth and Space Sciences, and
   Institute of Geophysics and Planetary Physics, University of
   California, Los Angeles, California 90095, USA. (werner@moho.ess.ucla.edu)}

\begin{document}

%
%
%
\setkeys{Gin}{draft=false}
%
%

%
%

\title{Constraints on the Size of the Smallest
Triggering Earthquake from the ETAS Model, B{\aa}th's Law, and
Observed Aftershock Sequences}
%

%
%


\author{Didier Sornette}
\affil{Department of Earth and Space Sciences, and Institute of
   Geophysics and Planetary Physics, University of California, Los
   Angeles, USA and Laboratoire de
   Physique de la Mati\`ere Condens\'ee, CNRS UMR6622, Universit\'e de
   Nice-Sophia Antipolis, France}

\author{Maximilian J. Werner}
\affil{Department of Earth and Space Sciences, and Institute of
   Geophysics and Planetary Physics, University of California, Los
   Angeles, USA}

\begin{abstract}
The physics of earthquake triggering together with
simple assumptions of self-similarity impose the existence
of a minimum magnitude $m_0$ below which earthquakes do not trigger other
earthquakes. Noting that the magnitude $m_d$ of completeness of seismic
catalogs has no reason to be the same as the magnitude $m_0$ of the
smallest triggering earthquake,
we use quantitative fits and maximum likelihood inversions of observed
aftershock sequences as well as B{\aa}th's law, compare with
ETAS model predictions and thereby constrain the value of $m_0$.
We show that the branching ratio $n$ (average number of triggered
earthquake per earthquake also equal to the fraction
of aftershocks in seismic catalogs) is the key parameter
controlling the minimum triggering magnitude $m_0$. Conversely,
physical upper bounds for $m_0$ derived from state- and
velocity-weakening friction
indicate that at least 60 to 70 percent of all earthquakes are aftershocks.

\end{abstract}

%
%

%

\begin{article}

%
%

\section{Introduction}
Scale invariance in earthquake phenomena is widely manifested
empirically, in the magnitude-frequency Gutenberg-Richter relation,
in the aftershock
Omori decay rate, and in many other relationships. Scale-invariance means
that there are no prefered length scales in seismogenic processes and in
spatio-temporal structures.
However, there are many reports that purport to identify characteristic scales.
As emphasized by {\it Matsu'ura} [1999], {\it Aki} [2000], and {\it
  Sornette} [2002], the search  
for characteristic structures in specific fault zones could allow the
separation of large earthquakes from small ones and thus advance
earthquake prediction.

Although there is clear evidence of deviations from self-similarity at large
scales [{\it Kagan}, 1999; {\it Pisarenko and Sornette}, 2003], the
issue is much murkier at small scales. For 
instance, {\it Iio} [1991], reports a lower magnitude cutoff $m_{\rm min} 
\approx -1.4$
for very small aftershocks of
the 1984 Western Nagano Prefecture, Japan, earthquake ($m_{\rm JMA} = 6.8$)
in spite of the fact that the high sensitivity of the observation system
(focal distances less than $1$~km and very low ground noise) would have
permitted to detect much smaller magnitudes.
Based on induced seismicity associated with deep gold mines,
{\it Richardson and Jordan} [1985] find a lower magnitude cutoff
$m_{\rm min} \approx 0$ 
for friction-dominated earthquakes, while
fracture-dominated earthquakes have no lower cutoff but an upper
cut-off of magnitude $\approx 1$. Using deep borehole
recordings, {\it Abercrombie} [1995a; 1995b] found that
small earthquakes exist down to at least magnitude $0$ and that source scaling
relationships hold down to at least $-1$. Based on
seismic power spectra, on the evidence
of a low-velocity low-Q zone reaching the top
of the ductile part of the crust and on
seismic guided waves in fault zones, {\it Li et al.} [1994] argue for a
characteristic
earthquake magnitude of about $3$ associated with the
width of fault zones. Another characteristic
magnitude in the range $4-5$ is proposed by {\it Aki} [1996], based on the
simultaneous change of coda $Q^{-1}$ and the fractional rate of
occurrence of earthquakes in this magnitude interval.
The existence of a discrete hierarchy of scales has in addition been suggested
by {\it Sornette and Sammis} [1995] based on the analysis of accelerated
seismicity prior to large earthquakes and recently by {\it Pisarenko
  et al.} [2004] 
by using a non-parametric measure of deviations from power laws
applied to the magnitude-frequency distributions of earthquakes in
subduction zones. Evidence of a hierarchy of scales is also found
in fragmentation and rupture processes
[{\it Sadovskii}, 1999; {\it Geilikman and Pisarenko}, 1989; {\it
    Sahimi and Arbabi}, 1996; {\it Ouillon et al.}, 1996; {\it
    Johanson and Sornette}, 1998; {\it Suteanu}, 2000].

    From a theoretical point of view,
the equation of motion for a continuum solid is scale-independent,
suggesting that deformation processes in solids should produce self-similar
patterns manifested in power law statistics. However, the symmetry
of the equation does not warrant that the solutions of this equation
share the same symmetry. Actually, the difference (when it exists) in
the symmetry
between a solution and its governing equation is known as the
phenomenon of ``spontaneous symmetry breaking'' [{\it Consoli and
    Stevenson}, 2000] 
underlying a large
variety of systems (for instance explaining the non-zero masses
of fundamental particles [{\it Englert}, 2004]). Of course, length scales
associated with
rheology and existing structures can produce deviations from exact
self-similarity. For instance, a transition
from stable creep to a dynamic instability at a nucleation size whose
dimensions depend on frictional and elastic parameters defines
a minimum earthquake size [{\it Dieterich}, 1992], estimated at magnitude
$\approx -3$
by {\it Ben-Zion} [2003]. This minimum size corresponds
only to events
triggered according to the mechanism of unstable sliding controlled
by slip weakening and thus concerns friction-dominated earthquakes.

A different perspective is offered by models of triggered seismicity
in which earthquakes
(so-called foreshocks and mainshocks) trigger other earthquakes
(so-called mainshocks and aftershocks, respectively). Recent studies
suggest that maybe more than 2/3 of events are
triggered by previous earthquakes (see {\it Helmstetter and Sornette}
[2003b] and 
references therein). In this context, the relevant question
is no more how small is the smallest earthquake but how small is the
smallest earthquake which
can trigger other earthquakes (and, in particular, larger earthquakes).

%
%


%
%

\section{The ETAS model and the smallest triggering earthquake}
To make this discussion precise, let us consider the
epidemic-type aftershock sequence (ETAS) model, in which any
earthquake may trigger other earthquakes, which in turn may trigger
more, and so on. Introduced in slightly
different forms by {\it Kagan and Knopoff} [1981] and {\it Ogata}
[1988], the model describes statistically the 
spatio-temporal clustering of seismicity.

The ETAS model consists of three assumed laws about the nature of seismicity
viewed as a marked point-process. We restrict this study to the temporal
domain only, summing over the whole spatial domain of interest.
First, the magnitude of any earthquake,
regardless of time, space or magnitude of the mother shock, is
drawn randomly from the exponential Gutenberg-Richter (GR) law.
Its normalized probability density function (pdf) is expressed as

\be
P(m) = {b \ln(10)10^{-b m} \over 10^{-b m_0}-10^{-b m_{max}}},
~~~~~m_0 \leq m \leq m_{max},
\label{GR}
\ee
where the exponent $b$ is typically close to one, and the cut-offs $m_0$
(see below) and $m_{max}$ serve to normalize the pdf. The upper 
cut-off $m_{max}$ is
introduced to avoid unphysical, infinitely large earthquakes. Its
value was estimated to be in the range $8-9.5$ [{\it Kagan}, 1999]. As the
impact of a finite $m_{max}$ is quite weak in the calculations below,
replacing the abrupt cut-off $m_{max}$ by a smooth taper
would introduce negligible corrections to our results.

Second, the model assumes that direct aftershocks are distributed in time
according to the modified ``direct'' Omori law (see {\it Utsu et al.} [1995]
and references therein). Assuming $\theta > 0$, the
normalized pdf of the Omori law can be written as
\be
\Psi(t) = {\theta c^{\theta}\over (t + c)^{1+\theta}} .
\label{pvalue}
\ee

Third, the number of direct aftershocks of an event of magnitude $m$
is assumed to follow the productivity law:
\be
\rho(m) = k ~10^{\alpha (m-m_0)},~~~~~  m_0 \leq m \leq m_{max}.
\label{formrho}
\ee

Note that the productivity law (\ref{formrho}) is zero below
the cut-off $m_0$, i.e. earthquakes smaller than $m_0$ do not trigger
other earthquakes, as is typically assumed in studies using the ETAS
model. The existence of the small-magnitude cut-off $m_0$ is
necessary to ensure the
convergence of the models of triggered seismicity (in statistical
physics of phase transitions and in particle physics, this is called
an ``ultra-violet'' cut-off which is often necessary to make the
theory convergent).
Below, we show that there are observable
consequences of the existence of the cut-off $m_0$ thus providing
constraints on its physical value.

The key parameter
of the model is defined as the number $n$ of direct
aftershocks per earthquake, averaged over all magnitudes. Here, we
must distinguish between the two cases $\alpha=b$ and $\alpha \neq b$:
\ba
     n & & \equiv \int \limits_{m_0}^{m_{max}} P(m) \rho(m) dm
     \nonumber \\
  && = \left\{ \begin{array} {r@{\quad \quad}l}
{k b \over b-\alpha} ({1-10^{-(b-\alpha)(m_{max}-m_0)} \over
    1-10^{-b(m_{max}-m_0)}}), & \alpha\neq b
\\ {k b \ln(10) (m_{max}-m_0) \over 1-10^{-b(m_{max}-m_0)}}, &
\alpha=b.\end{array} \right.
\label{nvaqlue}
\ea

Three regimes can be distinguished based on the value of $n$. The case
$n<1$ corresponds to the subcritical regime, where aftershock
sequences die out with probability one. The case $n>1$ describes
unbounded, explosive seismicity that may lead to finite time
singularities [{\it Sornette and Helmstetter}, 2002]. The critical
case $n=1$ separates the two regimes.

The fact that we use the same cut-off for the
productivity cut-off and the
Gutenberg-Richter (GR) cut-off is not a restriction as long as the real
cut-off for the Gutenberg-Richter law is smaller than or equal to the
cut-off for the productivity law. In that case, truncating the GR law
at the productivity cut-off just means that all smaller earthquakes,
which do not trigger any events, do not participate in the
cascade of triggered events. This should not be confused with
the standard incorrect procedure in many previous studies of 
triggered seismicity
of simply replacing the GR and productivity cut-off $m_0$ with the
detection threshold $m_d$ in equations (\ref{GR}) and (\ref{formrho})
(see, for
example, {\it Ogata} [1988]; {\it Zhuang et al.} [2004]; {\it
  Helmstetter and Sornette} [2002]; {\it Console et al.} [2003]; {\it
  Ogata} [1998]; {\it Felzer et al.} [2002]; {\it Kagan} [1991]).
This may lead to a bias in the estimated parameters.

The realization that the
detection threshold $m_d$ and the triggering threshold $m_0$
are different leads to
the question of whether we can extract the size of the smallest triggering
earthquake. Here, we infer useful information on $m_0$ from the physics
of earthquake triggering embodied in the simple ETAS formalism,
from B{\aa}th's law and from available catalogs.

There is no loss of generality in considering one (independent) branch
(sequence or cascade of aftershocks) of the ETAS model. Let an
independent background event of magnitude $M_1$ occur at some origin
of time. The mainshock will trigger direct aftershocks according to
the productivity law (\ref{formrho}). Each of the direct aftershocks
will trigger their own aftershocks, which in turn produce their own,
and so on.
Averaged over all magnitudes, each aftershock produces $n$ direct
offspring according to (\ref{nvaqlue}). Thus,
in infinite time, we can write the average of the total number
$N_{total}$ of direct and
indirect aftershocks of the initial mainshock as an infinite sum
over terms of (\ref{formrho}) multiplied by $n$ to the power of the
generation [{\it Helmstetter et al.}, 2004], which can be expressed
for $n<1$ as: 
\ba
N_{total} & &  = \rho(M_1)+\rho(M_1) n+\rho(M_1) n^2 + ... \nonumber \\
&& = {k ~10^{\alpha (M_1-m_0)} \over 1-n}
\label{Ninf}
\ea
However, since we can only detect events above the detection threshold $m_d$,
the total number of observed aftershocks $N_{obs}$ of the sequence
is simply $N_{total}$ multiplied by the fraction of events above
the detection threshold, given by
$(10^{-b(m_{max}-m_d)}-1)(10^{-b(m_{max}-m_d)}-1)^{-1}$ according
to the GR distribution. The observed number of events in the sequence
is therefore
\ba
N_{obs} & =& N_{total} ~\left({10^{b(m_{max}-m_d)}-1 \over
10^{b(m_{max}-m_0)}-1}\right) \nonumber \\
& = &  {k ~~10^{\alpha (M_1-m_0)} \over  1-n}~
\left({10^{b(m_{max}-m_d)}-1 \over 10^{b(m_{max}-m_0)}-1}\right)~.
\label{Nobs}
\ea

Equation (\ref{Nobs}) predicts the average observed number of direct and
indirect aftershocks of a mainshock of magnitude $M_1>m_d$. To
estimate $m_0$, we need to eliminate or find estimates of the three
unknowns $n$, $k$, and $N_{obs}$. We can eliminate $k$ through the
expression (\ref{nvaqlue}) for $n$, leaving $n$ and $N_{obs}$. The
mean number of observed aftershocks as a function of mainshock
magnitude $M_1$ was estimated by {\it Helmstetter et al.} [2004] and
{\it Felzer et al.} [2003] and can also be obtained from B{\aa}th's law. In
the following sections, we use these three estimates for $N_{obs}$ and
thus obtain $m_0$ as a function of the only remaining unknown
$n$. Acknowledging the controversy surrounding the estimation of the
percentage of aftershocks in a catalog, we nevertheless use existing
estimates of $n$ to finally obtain quantitative values for $m_0$.

\section{Constraint on the smallest triggering earthquake from the
      ETAS model and observed estimates of aftershock numbers}

Following the recipe outlined above, we begin by using the estimates
of the observed number of aftershocks $N_{obs}$ obtained by
{\it Helmstetter et al.} [2004] in order to find $m_0$ as a function of
$n$. {\it Helmstetter et al.} [2004] sidestepped the problems associated
with maximum likelihood estimates of the complete model parameters by
fitting stacked observed aftershock rates within pre-defined
space-time windows using the formula

\be
\lambda_{fit}(t) = {K_{fit} ~~10^{\alpha M_1 - b m_d} \over t^{p_{fit}}}~,
\label{lambda_fit}
\ee
based on scaling laws (the GR law, the Omori law, and
the productivity law) discussed above.
The constant $K_{fit}$ includes all aftershocks, direct and indirect,
and thus corresponds to a global renormalized constant different from
$k$ in the ETAS productivity law (\ref{formrho}). Furthermore,
$p_{fit}$ is also a global
exponent, which may be different from the local exponent $1+\theta$ of
the ETAS model for $n$ close to $1$ and at not too long times, as
explained in {\it Sornette and Sornette} [1999] and {\it Helmstetter
  and Sornette} [2002]. The total number of aftershocks 
is then obtained by integrating over an
un-normalized Omori law according to {\it Helmstetter et al.} [2004]:
\ba
N_{fit}(T,M_1) & &= \int \limits_c^T \lambda_{fit}(t) dt \nonumber \\
&& = K_{fit} ~10^{\alpha M_1 - b m_d} ~{T^{1-p_{fit}}-c^{1-p_{fit}} 
\over 1-p_{fit}}.
\label{N_fit_T}
\ea
For $p_{fit}<1$, this expression diverges as $T$ increases to infinity. But, as
it has been shown that the exponent of the observed, global Omori law
converges to a value $1+\theta >1$ at large times for $n<1$ [{\it
  Sornette and Sornette}, 1999; {\it Helmstetter and Sornette}, 2002], 
the time factor converges also to $(\theta c^{\theta})^{-1}$.
Under the assumption that $p_{fit}=1+\theta>1$, valid for $n$ not very
close to the critical value $1$, equation (\ref{N_fit_T}) may then be
rewritten as
\be
N_{fit}(M_1)= K_{fit}~ 10^{\alpha M_1 -  b m_d} ~(\theta c^{\theta})^{-1}.
\label{N_fit}
\ee
Equating the ETAS model prediction $N_{\rm obs}(M_1)$ given by
(\ref{Nobs}) with the empirical estimate $N_{\rm fit}(M_1)$ given by
(\ref{N_fit}), and eliminating the unknown $k$ through the expression
for $n$ in (\ref{nvaqlue}) leads to an equation for $m_0$ as a
function of $n$:
\be
m_0   =  {1 \over (b-\alpha) \ln(10)} \times    \nonumber
\ee
\be 
\ln (10^{(\alpha-b)m_{max}}
   +  {b - \alpha \over b} {n \over 1-n}	{\theta c^{\theta} \over
	 K_{fit}} (1-10^{-b(m_{max}-m_d)}))~, 
	 \label{mglre}
\ee
for $\alpha \neq b$ and 
\be
m_0   = m_{max}  - ({n \over 1-n}) {\theta c^{\theta} \over
    K_{fit}} {1-10^{-b(m_{max}-m_d)} \over b \ln(10)}, ~,
\ee
for $\alpha = b$.
Expression (\ref{mglre}) shows that, provided an estimate of
the branching ratio $n$ is available,
we can deduce $m_0$, since the other quantities
can be measured independently: $b$ is close to 1, $\alpha$ is usually between
0.5 and 1, $m_d$ depends on catalogs but is often about $3$, $c$ is
typically close to $0.001$ days and $K_{fit}$ given
by (\ref{lambda_fit}) is obtained from the calibration of the
productivity of earthquakes as a function of their magnitude. In Table
1 of their study, {\it Helmstetter et al.} [2004] report values for
$K_{fit}$ in the range from $0.005$ to $0.02$ (days)$^{p-1}$, $0.94
\leq \alpha \leq 1.05$, $b \approx 0.95$, $m_d=[2, 3, 4]$, $c=0.001$
days and $\theta=0.1$.

We note that $m_d$ appears in the expression (\ref{mglre}) for
$m_0$. Clearly, a detection threshold that evolves with seismic
technology should not influence the physics of triggering. We thus
expect $m_0$ to be independent
of $m_d$. The reason $m_d$ does appear in the expression can be traced
to the GR law (\ref{GR}), which is normalized over the magnitude interval from
$m_0$ to $m_{max}$. When integrated to give the probability of $m$
lying in the range from $m_d$ to $m_{max}$, the factor involving $m_d$ does
not enter as simply as in the formulation (\ref{lambda_fit}) of
{\it Helmstetter et al.} [2004]. Therefore the factors do
not cancel out when comparing the ETAS prediction with the assumed
parameterization of {\it Helmstetter et al.} [2004]
and $m_d$ remains in the equations. Assuming that the GR law is
correctly normalized in the present ETAS model, this implies a (weak)
dependence of $K_{fit}$ on $m_d$. Given the correlation between
$\alpha$ and $K_{fit}$ (see below), the estimates of $\alpha$ may thus
also depend on $m_d$.

\begin{figure}
\noindent\includegraphics[width=40pc]{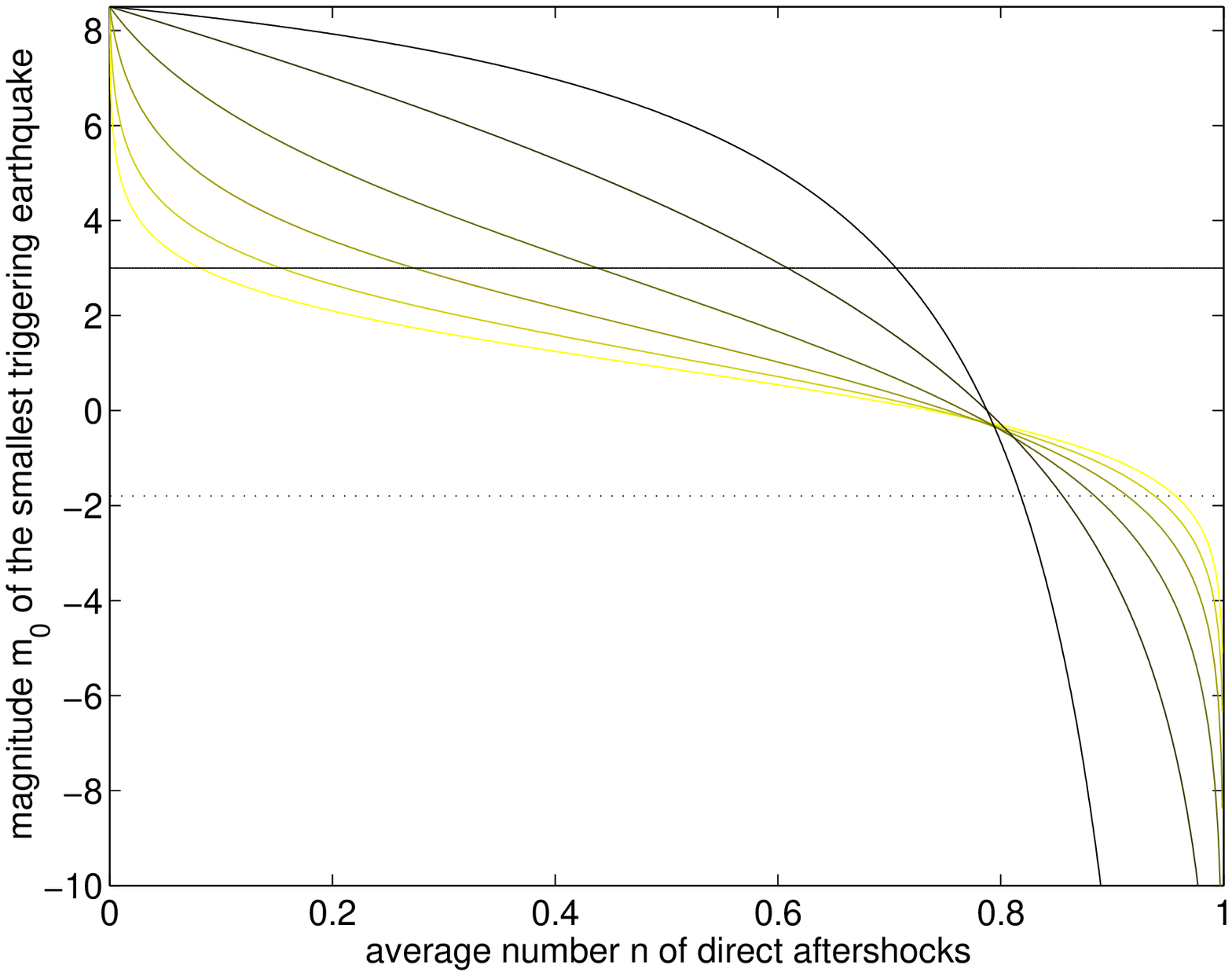}
  \caption{The magnitude $m_0$ of the smallest triggering earthquake as a
        function of the average number $n$ of direct aftershocks
        estimated from fits to observed aftershock sequences and the
        ETAS model for values of, from light to dark,  $[\alpha=0.5,
        K_{fit}=0.070]$, $[\alpha=0.6, K_{fit}=0.058]$, $[\alpha=0.7,
        K_{fit}=0.046]$, $[\alpha=0.8, K_{fit}=0.034]$, $[\alpha=0.9,
        K_{fit}=0.022]$, $[\alpha=b=1, K_{fit}=0.01]$. Common parameters
        are $b=1$, $m_{max}=8.5$, $\theta=0.1$, $c=0.001$
        days. $K_{fit}$, in units of
        days$^{1-p}$, was estimated by {\it Helmstetter et al.} [2004] and
        herein adapted to these values of $\alpha$ through their
        correlation (see Figure \ref{Correlation}). The horizontal lines
        represent upper limits
        of $m_0$, derived from the detection threshold $m_d=3$ (solid),
        and from estimates of the critical slip $D_c~1m$ in rate and state
        friction giving $m=-1.8$ (dashed), providing conversely a lower
        bound for the percentage of aftershocks in an earthquake catalog
        and/or $\alpha$.}
\label{m0_n}
\end{figure}

The estimate of $m_0$ that we are trying to obtain relies on
the adequacy of the model used here and on the stability and reliability
of the quoted parameters. For now, we sidestep any possible
difficulties in the determination of the parameters and
present in Figure \ref{m0_n} the magnitude of the smallest triggering
earthquake $m_0$ as a function of
the average number $n$ of direct aftershocks per mainshock for a
range of parameters. For $n=0$,
$m_0$ equals the largest possible
earthquake $m_{max}$, representing the limit that earthquakes do not
trigger any aftershocks. At the other end, for $n=1$, the formula
predicts that $m_0$ diverges to minus infinity. Recall that $n=1$
corresponds to the system being exactly at the critical value of a
branching process and the statistical average $N_{\rm obs}(m)$ of the
total number of events triggered over all generations by a mother
event of magnitude $m$ becomes infinite. Of course, individual
sequences have a finite lifetime and a finite progeny with probability
one and the theoretical average loses its meaning due to the fat-tailed
nature of the corresponding distribution [{\it Athreya}, 1972; {\it
    Saichev et al.}, 2004; {\it Saichev and Sornette}, 2004].
Therefore, the prediction on $m_0$ becomes unreliable for $n$ close to
$1$.

For a wide range of $n$ and combinations between $\alpha$ and $K_{fit}$,
the magnitude of the smallest triggering earthquake lies between $0$ and
$-10$. Only for values of $n$ above $0.9$ does the size of $m_0$
become smaller than $-10$. For reference, a magnitude $-10$ event corresponds
to a fault of length $1$mm, i.e. to a typical grain size.

Given that we expect $m_0$ to be smaller than the detection threshold 
$m_d$, the
horizontal line at $m_d=3$ serves as a (very) conservative estimate of
the upper limit of $m_0$ and thus provides constraints on the combination of
parameters $\alpha$, $K_{fit}$ and $n$. For example, for $\alpha=0.5$ and
$K_{fit}=0.0702$, at least 10 percent of all earthquakes must be
aftershocks. This lower limit increases drastically to about 70 percent for
$\alpha=b=1$ and $K_{fit}=0.0095$.

We can obtain another external bound on $m_0$ from estimates
of the minimum slip required before static friction drops to kinetic
friction and unstable sliding begins, according to models of
velocity-weakening friction. For example, the parameter $D_c$ in rate
and state dependent friction [{\it Dieterich}, 1992; {\it Dieterich},
  1994] was estimated at $0.5$ m from seismograms [{\it Ide and
      Takeo}, 1997] and similarly 
at $40-90$ cm from slip-velocity records [{\it Mikumo et al.}, 2003],
although both probably correspond to upper bounds. Estimates of $D_c$ 
from laboratory
friction experiments are
approximately 5 orders of magnitude less than the upper bound
determined by seismic studies. One could conclude that either the
upper bound from seismic studies is so extreme as to render the
comparison to laboratory studies meaningless, or the slip weakening
process is in fact different at laboratory scales
[{\it Kanamori and Brodsky}, 2004]. If we assume that the minimum slip
needed to 
initiate stable sliding corresponds to the minimum length of a
friction-based earthquake, then, neglecting fracture-based
earthquakes, $D_c$ corresponds to the size of the smallest
earthquake. Given that the smallest triggering earthquake
must be equal to or larger than the smallest earthquake, but that the
estimates of $D_c$ are upper limits, we use these values for $D_c$ as an
upper limit of the smallest triggering earthquake. From the
relations between fault length, moment and moment magnitude
[{\it Kanamori and Brodsky}, 2004] with $D_c=1$ m and a stress drop of
$3$ MPa, we 
obtain an upper limit of magnitude $-1.8$ for the smallest triggering
earthquake. This upper limit is respresented in Figure \ref{m0_n} as
the lower horizontal line.

\begin{figure}
\noindent\includegraphics[width=40pc]{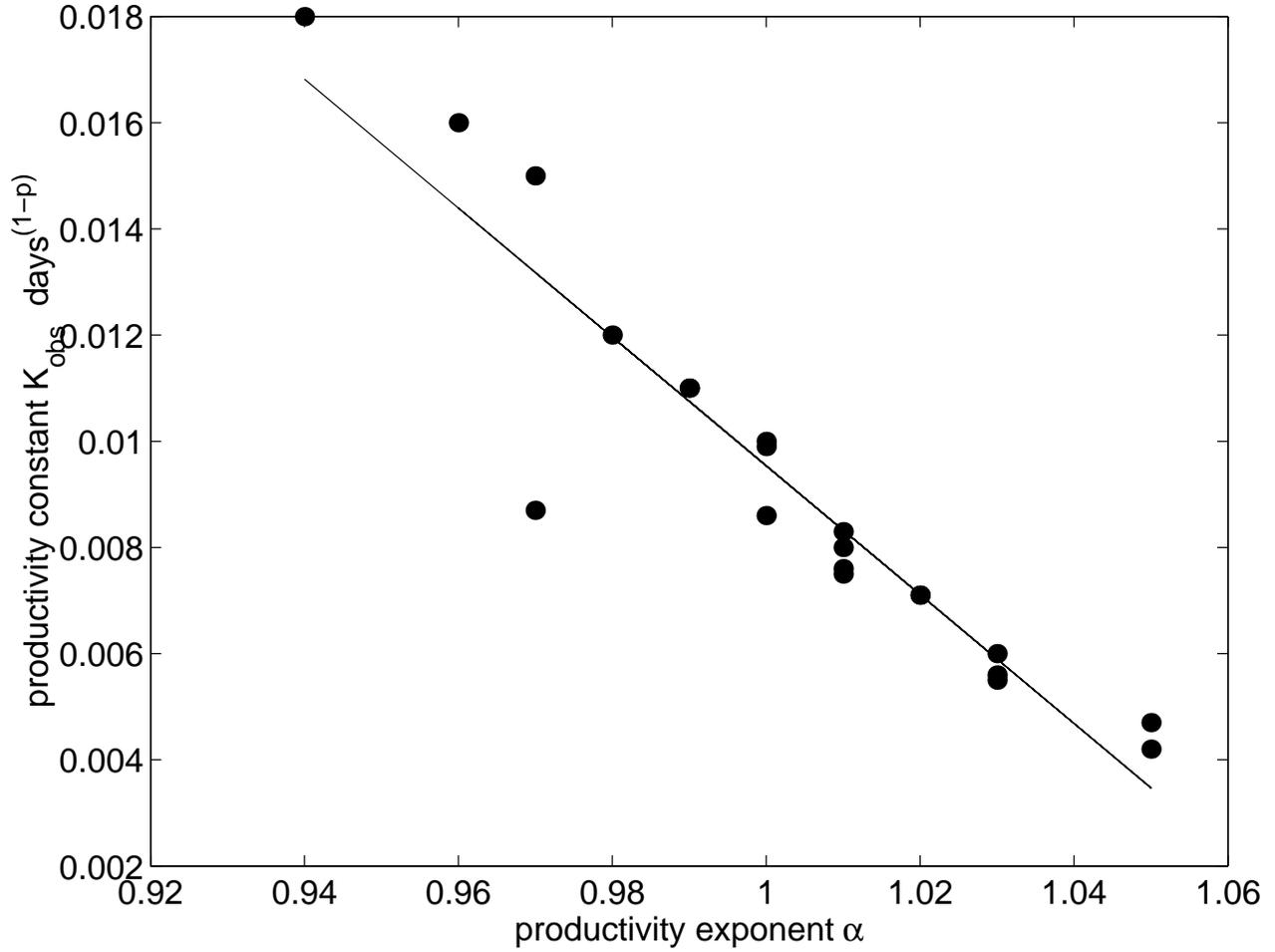}
        \caption{The correlation between the values of $K_{fit}$ and $\alpha$
        taken from Table 1 of {\it Helmstetter et al.} [2004]. The line is a 
least-squares
        fit with slope -0.1214 and y-intercept 0.1309. The extrapolation
        of this fit for smaller values of $\alpha$ was used to obtain the
        values for $K_{fit}$ in Figure \ref{m0_n}.}
\label{Correlation}
\end{figure}

{\it Felzer et al.} [2002] have
used $\alpha=b$ on the basis of an argument of self-similarity.
{\it Helmstetter et al.} [2004] also argue for a value of
$\alpha$ essentially undistinguishable from $b$ based on fits
of stacked aftershock decay rates in pre-defined space-time windows.
Other studies have found $\alpha$ as
small as and even smaller than $0.5$
(see, for example, {\it Console et al.} [2003]; {\it Helmstetter}
[2003]; {\it Zhuang et al.} [2004]). In
view of the lack of consensus and to keep the discussion independent
of the estimation problem, we use the correlation we found between the
parameters $K_{fit}$ and $\alpha$ estimated in
{\it Helmstetter et al.} [2004] to
extrapolate to smaller $\alpha$. The existence of such a correlation is
standard in joint
estimations of several parameters and can be deduced from the inverse of the
Fisher matrix of the log-likelihood function {\it Rao} [1965]. Such
correlation can 
also be enhanced if the model is misspecified. We performed a
least-square fit to
the scatter plot (see Figure \ref{Correlation}) to obtain a relationship
between the parameters and then calculated an estimate of $K_{fit}$
for smaller values of $\alpha$. The resulting curves are plotted in
Figure \ref{m0_n}.


Delaying the discussion on the estimation problem until the end of the
section, we use (\ref{mglre}) together with existing estimates of the
percentage of aftershocks in seismic catalogs (equivalent to $n$
[{\it Helmstetter and Sornette}, 2003b]) to constrain $m_0$. We note,
however, that different declustering techniques lead to different
estimates. No consensus exists on which method should
be trusted most. For example, {\it Gardner and Knopoff} [1974] found
that about 2/3 of the events 
in the Southern California catalogue are aftershocks. With
another method, {\it Reasenberg} [1985] found that 48\% of the events
belong to a 
seismic cluster. {\it Davis and Frohlich} [1991] used the ISC catalog and,
out of 47500 earthquakes, found that 30\% belong to a cluster,
of which 76\% are aftershocks and 24\% are foreshocks. Recently, using
different versions of the ETAS model, {\it Zhuang et al.} [2004] have
performed a careful inversion of
the JMA catalog for Japan using a magnitude $m_d=4.2$ for the
completeness of their catalog. They provide three estimates of the branching
ratio for their best model: $n=0.42$, $0.55$, and $0.46$.

Whether any of these methods estimate $n$ correctly and without
bias remains questionable. In particular, the
branching ratios as calculated by {\it Zhuang et al.} [2004] and
others may be significantly biased by the assumption that
$m_d=m_0$, which can be shown to lead to an apparent
branching ratio modified by the impact of
hidden seismicity below the catalog completeness [{\it Sornette and
    Werner}, 2004]. Moreover, there are problems with the maximum likelihood
estimation (see for instance {\it Helmstetter et al.} [2004]).
However, in the absence of better estimations, we
nevertheless use the above values as rough estimates of $n$.
Given the range of $\alpha$ and $K_{fit}$, $m_0$ is still not
very well constrained for one particular value of $n$ (see Figure
\ref{m0_n}). For example,
for 50 percent aftershocks ($n=0.5$), $m_0$ ranges from $1$ to an
unrealistic
$7$ depending on the values of $\alpha$ and $K_{fit}$. This argument
could be used to rule out the combination $n=0.5$ and $\alpha=1$. In
fact, for $m_0$ to be smaller than $m_d=3$, at least 70 percent of
earthquakes are aftershocks. For $m_0$ to be smaller than the upper
limit estimated from $D_c$, at least 80 percent must be aftershocks.

We can also use the values obtained by {\it Felzer et al.} [2003]
to constrain $m_0$. The authors also used finite space-time windows in
which they fitted aftershock sequences with parameters for a global
sequence according to
\be
C_T={A_T \over 1-p_T} ( (t+c_T)^{1-p_T}-{c_T}^{1-p_T}), ~~~p_T \neq 1,
\label{Nf_t}
\ee
where $t$ is the selected duration of the sequences, $p_T$ is the
global Omori exponent, $c_T$
is the Omori constant, and $A_T$ is the productivity. Assuming
that the local Omori exponent is $p=p_T=1+\theta_T$, expression (\ref{Nf_t})
can be rewritten for the infinite time limit as
\be
C_T={A_T \over \theta_T {c_T}^{\theta_T}}.
\label{Nf}
\ee
The obtained values are listed in Table 3 of their study: $A_T=0.116$
days$^{1-p_T}$, $p_T=1.08$ and $c_T=0.014$ days. These values hold for
a typical California aftershock sequence of a magnitude $M_1=6.04$
mainshock, a detection threshold of $m_d=4.8$, and $\alpha=b=1$.

\begin{figure}
\noindent\includegraphics[width=40pc]{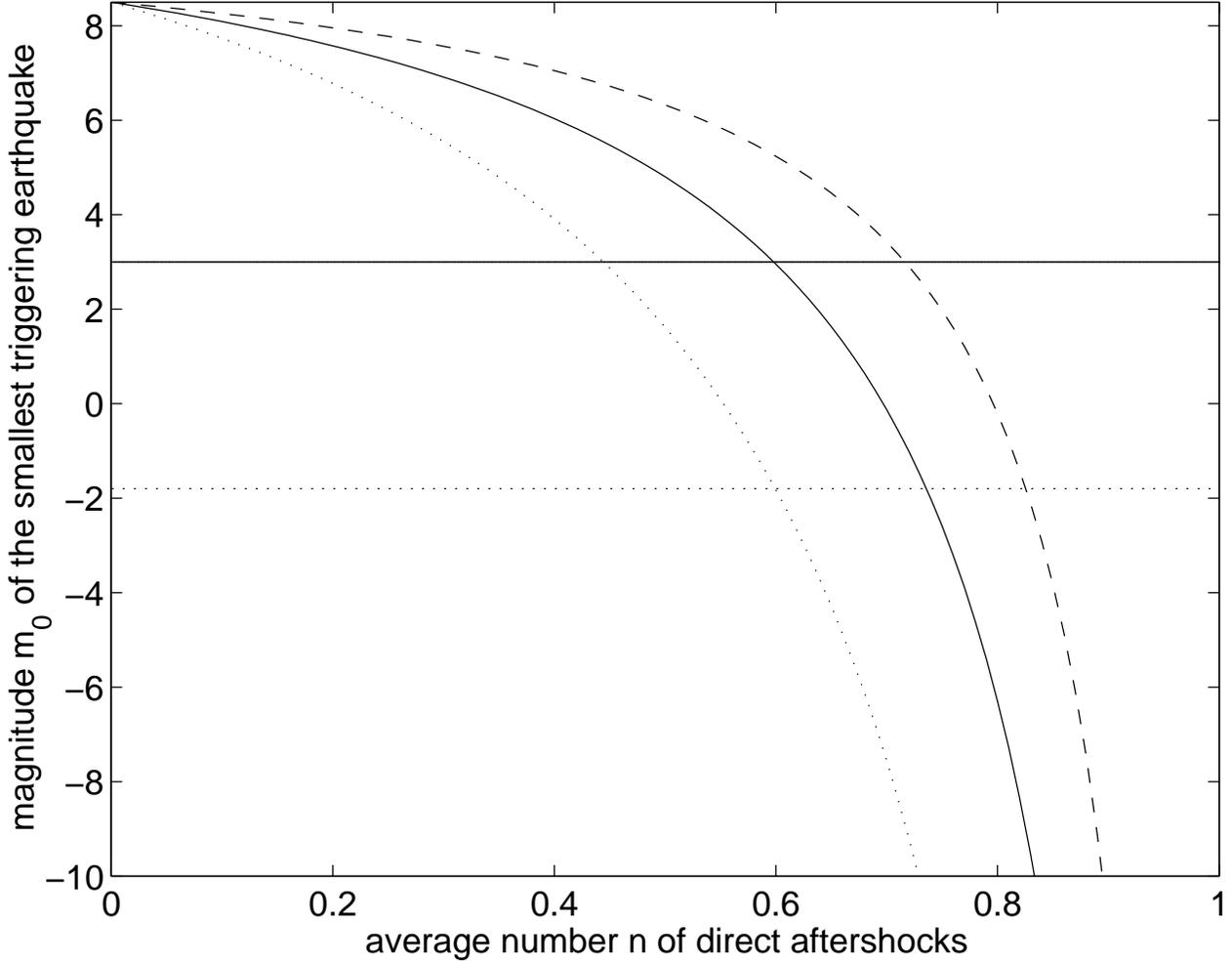}
        \caption{The magnitude $m_0$ of the smallest triggering earthquake as a
        function of the average number $n$ of direct aftershocks
        estimated from fits to observed aftershock sequences and the
        ETAS model according to {\it Felzer et al.} [2003] (solid),
        using the parameters $\alpha=b=1$, $m_{max}=8.5$,
        $\theta_T=0.08$, $c_T=0.014$ days, $M_1=6.04$ and $A_T=0.116$
        days$^{1-p_T}$. For comparison, we include the curves
        corresponding to the special case $\alpha=b$ for the fit
        according to {\it Helmstetter et al.} [2004]
        (dashed) and the constraint due to B{\aa}th's law (dotted) (see
        \ref{m0_bath}). The horizontal lines represent upper limits
        of $m_0$, derived from the detection threshold $m_d=3$ (solid),
        and from estimates of the critical slip $D_c~1m$ in rate and state
        friction giving $m=-1.8$ (dashed), providing conversely a lower
        bound for the percentage of aftershocks in an earthquake catalog. }
\label{m0_felzer}
\end{figure}

As before,  we equate the ETAS model prediction (\ref{Nobs}) with the
observation (\ref{Nf}), eliminate $k$ through expression
(\ref{nvaqlue}) for $n$ (where $\alpha=b$) and obtain an equation for
$m_0$ as a function of $n$:
\ba
m_0 & =& m_{max} -{n \over 1-n} {(1-10^{-b(m_{max}-m_d)})\over b
   \ln(10)} \nonumber \\
  && \times {\theta_T c^{\theta_T} \over A_T} 10^{b(M_1-m_d)}, ~~~~~ 
\alpha=b=1.
\label{m0f}
\ea
This expression for $m_0$ is shown in Figure \ref{m0_felzer} (solid
curve). As in equation (\ref{mglre}), $m_d$ remains artifactually in
the equation due to a
dependence of $A_T$ on the detection threshold. Since the
parameters were obtained with $\alpha=b=1$, we do not alter the values
of $\alpha$ and obtain only one curve. For comparison, we include the
curve for the case $\alpha=b$ that we obtained above in Figure \ref{m0_n},
based on the fits
by {\it Helmstetter et al.} [2004] (dashed curve) and the
curve for the case $\alpha=b$ that results from using B{\aa}th's law
(see next section) (dotted). We observe the same characteristics as
before in that $m_0$ approaches $m_{max}$ for small $n$ and that it
diverges to minus infinity for $n$ going to one. Differences between 
the three curves
arise only in the faster or slower decrease of $m_0$ with $n$. For
example, the (conservative) upper limit $m_d=3$ for $m_0$ constrains $n$
to be larger than 60 percent according to the values obtained by
{\it Felzer et al.} [2003], whereas the parameters of {\it Helmstetter
  et al.} [2004] for the case $\alpha=b$ impose $n$ to be at least 70
percent. For the estimate obtained from B{\aa}th's law (see
next section), $n$ must be larger than about 45 percent. If we assume
that the upper limit of $m_0$ can be obtained from estimates of $D_c$,
corresponding to $m=-1.8$, then $n$ must be at least 60 percent
according to the estimate from
B{\aa}th's law, 75 percent according to {\it Felzer et al.} [2003],
and 85 percent according to the fit by {\it Helmstetter et al.}
[2004]. Conversely, $n=0.7$ determines $m_0$ roughly equal to zero,
whereas $n=0.8$ implies $m_0$ lies in the range $-7$ to $-5$.

Since the three expressions for $m_0$ (for $\alpha=b$) show the
same functional dependence on key variables and differ only in the
different estimates of a few constants, this consistency provides some
confidence in our results. As for the difference in the three
curves, they constitute three differently formulated, empirical
estimates of the number of events of
typical aftershock sequences. Given the variability of the
aftershock process, the discrepancy in the estimates is to be expected.

We now point out difficulties for exploiting quantitatively the
above ideas. Our conclusions for $m_0$ and $n$ are based on empirical
parameter estimations that involve delicate technical problems.
The constants $K_{\rm obs}$ defined in (\ref{lambda_fit}) and $A_T$
defined in (\ref{Nf}) are in principle measurable. Many issues may
bias the estimation of these parameters: (i) The total number of
aftershocks is estimated empirically in finite space and time windows
and events outside are thus missed. In particular, in the special case
where $p$, as defined in (\ref{pvalue}), and $n$ are both close to
$1$, a substantial fraction of aftershocks occurs at very long time
and are very difficult if not impossible to distinguish from the
background seismicity. (ii) Stacking different sequences with
different global Omori law decays may introduce errors. (iii) The $p$
exponent of the Omori law may intrinsically depend on the mainshock
magnitude [{\it Ouillon and Sornette}, 2004]. (iv)
Background events may be falsely counted as aftershocks. (v)
Magnitude and location uncertainties may bias the
parameters. (vi) Missing events in the catalog, especially after
large events, may artifactually change the parameter values. (vii)
Undetected seismicity may bias the estimated parameters [{\it Sornette
    and Werner}, 2004].

\section{Constraints on the smallest triggering earthquake from Bath's
      law}

Finally, we use the empirical B{\aa}th's law to constrain $m_0$ as a
function of $n$. The law states that the average difference between a
mainshock of magnitude $M_1$ and the magnitude $m_a$ of
its largest aftershock is $dm=M_1-m_a= 1.2$, regardless of the
mainshock magnitude (see for example {\it Helmstetter and Sornette},
  [2003a] and references therein).
Let $N_{obs}$ be the total number of aftershocks generated
by the mainshock above the magnitude $m_d$ of completeness of the
catalog. Assuming that the magnitudes of the aftershocks are drawn from the
Gutenberg-Richter law, the largest aftershock has an average magnitude given by
a simple argument of extreme value theory:
\be
m_a = m_d+(1/b) (\log_{10} N_{obs})~.
\label{m_a}
\ee
Solving this expression for $N_{obs}$, equating it with the
ETAS prediction (\ref{Nobs}) and eliminating $k$ through
the expression for $n$ (\ref{nvaqlue}) provides an estimate of $m_0$
as a function of $n$:
$$
m_0= {1 \over (\alpha-b) \ln(10)}  \times
$$
\be
 \ln 10^{(\alpha-b)m_{max}} +
         {b-\alpha \over b} {n \over 1-n}~10^{\alpha M_1 -b m_a}
         ~(1-10^{-b(m_{max}-m_d)}), 
         \label{m0bath}
\ee
for $\alpha \neq b$ and
\be
m_0= m_{max}-({n \over 1-n}) {(1-10^{-b(m_{max}-m_d)}) \over b
    \ln(10)}~10^{b(M_1-m_a)}~,
\ee
for $\alpha = b$.

\begin{figure}
\noindent\includegraphics[width=40pc]{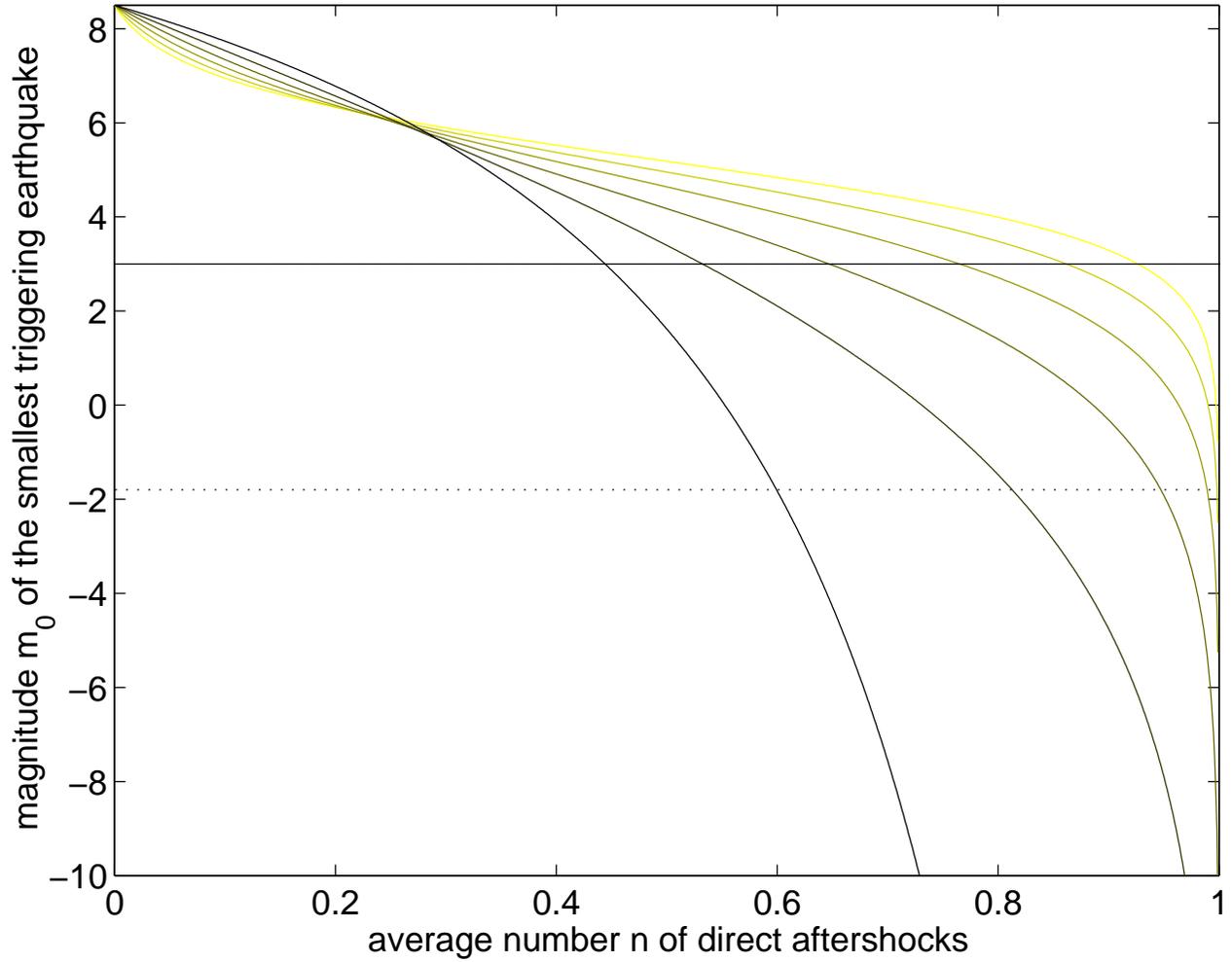}
        \caption{The magnitude $m_0$ of the smallest triggering
        earthquake as a function of the average number $n$ of direct
        aftershocks as estimated by the ETAS model and B{\aa}th's law
        for $\alpha=[0.5, 0.6, 0.7, 0.8, 0.9,
        1]$, from light to dark. The horizontal lines represent upper limits
        of $m_0$, derived from the detection threshold $m_d=3$ (solid),
        and from estimates of the critical slip $D_c~1m$ in rate and state
        friction giving $m=-1.8$ (dashed), providing conversely a lower
        bound for the
        fraction $n$ of aftershocks in an earthquake catalog and/or $\alpha$.
        Common parameters are $b=1$, $m_{max}=8.5$, $m_{main}=7$, $m_a=5.8$.}
\label{m0_bath}
\end{figure}

Figure \ref{m0_bath} illustrates the behaviour of $m_0$ as a function of
the average number $n$ of direct aftershocks for reasonable values of the
other constants ($m_{max}=8.5$, $m_d=3$, $b=1$, $\alpha=[0.5, 0.6, 0.7, 0.8,
    0.9, 1]$ (light to dark)), for mainshock and largest aftershock
values according to $M_1-m_a=7-5.8=1.2$ from B{\aa}th's law. Again, as
$n$ tends to one, $m_0$ tends to minus infinity, while for $n=0$,
$m_0=m_{max}$, as expected. We also observe that $m_0$ is almost
constant over a wide range of $n$ for comparatively small $\alpha$,
whereas $m_0$ varies much faster for the case $\alpha=b$.

As alluded to in the last section, we obtain the same functional
dependence as in both previous estimates
of the last section. However, for
$\alpha=b$, the decrease of $m_0$ with increasing $n$ is even faster
than when using the parameters of {\it Felzer et al.} [2003]. Here, the
upper limit $m_d=3$ for $m_0$ (upper horizontal line) constrains $n$ to be
larger than 45 percent, smaller than the 60 percent found
previously. This discrepancy
is due to the three different ways of estimating the observed number
of aftershocks. However, since all three are in a similar range, they
provide a test of the consistency of the results.

When applying the $D_c$-derived upper limit of $-1.8$ (lower
horizontal line), $n$ must be larger than at least 60 percent for
$\alpha=b$ and larger than 80 percent for $\alpha=0.9$. We thus find that
for $n=0.5$, $m_0$ is in the range $2$ to an unrealistic $5$, while
for $n=0.7$, $m_0$ lies between $-10$ and $5$, depending on the values
of $\alpha$. Since $m_0 \geq 3$ is unrealistic, the entire region of
combinations between $\alpha$ and $n$ that fall above that value can
be ruled out. For example, the case $\alpha=0.8$ leads to a reasonable
$m_0$ smaller than $m_d=3$ only for $n$ larger than about 65 percent.

\section{Conclusions}

We have shown that differentiating between the smallest triggering
earthquake $m_0$ and the detection threshold $m_d$ within the ETAS
model leads, together with three separate estimates of the
observed numbers of aftershocks, to three estimates of $m_0$ as a
function of the percentage $n$ of aftershocks in a catalog
(also equal to the branching ratio). We have used
empirically fitted values for aftershock numbers and thereby
eliminated one variable from the ETAS formalism in order to obtain an
estimate of $m_0$ as a function of $n$. The three
different estimates were obtained from the fits performed by
{\it Helmstetter et al.} [2004], by {\it Felzer et al.} [2003], and from the
empirical B{\aa}th's law (see {\it Helmstetter and Sornette} [2003a]
and references therein). All three give
the same functional dependence and similar values for $m_0$. In
particular, we can place bounds on $m_0$ from estimates of the
percentage of aftershocks in earthquake catalogs. Conversely, we can
limit the range of $n$ by observing that $m_0$ must be less than the
detection threshold $m_d$, or, less conservatively, that $m_0$ must be less
than the magnitude corresponding to the rate-and-state critical slip
$D_c$ as estimated from seismograms. Apart from quantitative values
for $m_0$, the bounds limit the possible combinations between $n$ and
$\alpha$ and, in particular, indicates that at least 60 to 70 percent
of all earthquakes are aftershocks.

The fact that the existence of a small magnitude cut-off $m_0$ for triggering
should have observable consequences may appear surprising. But such a
phenomenon of the impact of a small scale cut-off
on ``macroscopic'' observables is not new in physics and actually
permeates particle physics, field theory and condensed matter
physics. In the present case, the existence of $m_0$ has an observable
impact especially when $\alpha \leq b$ for which the cumulative effect
of tiny earthquakes dominate or equate that of large earthquakes with
respect to the physics of triggering other earthquakes
[ {\it Helmstetter}, 2003; {\it Helmstetter et al.}, 2004]. We hope
that the present article, together with our companion paper
{\it Sornette and Werner} [2004], will draw the 
attention of the community to the important problem of the distinction
between $m_d$ and $m_0$. Moreover, it will perhaps encourages re-analyses of
inversion methods of models of triggered seismicity, and in particular
of maximum likelihood estimations, to take into account the bias due
to the unobserved seismicity below the magnitude of catalog completeness.

\begin{acknowledgments}
We acknowledge useful discussions with
A. Helmstetter, K. Felzer and J. Zhuang.
This work is partially supported by NSF-EAR02-30429, by
the Southern California Earthquake Center (SCEC).
SCEC is funded by NSF Cooperative Agreement EAR-0106924 and USGS Cooperative
Agreement 02HQAG0008.  The SCEC contribution number for this paper is xxx.
\end{acknowledgments}

%
%
%
%
%
%
%
%


\end{article}

\pagebreak

%
%
%

\pagebreak

\pagebreak

\pagebreak

\end{document}